# Van Vleck paramagnetism and enhancement of effective moment with magnetic field in rare earth orthovanadate EuVO$_4$

Dheeraj Ranaut and K. Mukherjee*

School of Physical Sciences, Indian Institute of Technology Mandi, Mandi 175075, Himachal Pradesh, India

Email: kaustav@iitmandi.ac.in

**Abstract**

The *4f*$^6$ systems were hypothesized to possess non-magnetic ($J = 0$) ground state. However, all such systems have distinctly shown the presence of non-zero effective moment. In this context, a rare earth orthovanadate EuVO$_4$, which in spite of having $J = 0$ ground state, possess non-zero magnetic moment. Our studies reveal three different regions in this compound, boundaries of which are demarcated from susceptibility data. The high temperature susceptibility exhibits linear dependence on temperature which arises due to tetragonal crystal field, followed by a temperature independent plateau like region, ascribed to Van Vleck paramagnetism. At low temperatures, Curie-Weiss like behaviour is observed, which arise due to magnetic Eu$^{2+}$ moments and results in non-zero effective moment. Our analysis reveals that the separation ($\lambda$) between the $J = 0$ and $J = 1$ states decreases on increasing the external magnetic field which leads to an enhanced effective moment at higher fields.

# 1. Introduction:

Rare earth-based systems have attracted significant interest in the field of magnetism due to the presence of partially filled $4f$ orbitals. In this family, $Eu^{3+}$ based compounds exhibit magnetic properties which are different from other members [1-5]. Since $Eu^{3+}$ has six $4f$ electrons, according to Hund's rule, it has same value of orbital and spin angular momentum ($L = S = 3$). This results in total angular momentum $J = 0$, and thus, hypothetically the ground state should be non-magnetic ($^7F_0$). For most of the rare earth materials, the magnetic susceptibility follows the conventional Curie's law. But the exception lies for two rare earth elements, $Eu^{3+}$ and $Sm^{3+}$, which have very small energy separation between the ground state and excited states [1]. For $Eu^{3+}$ compounds, the separation between the ground state $^7F_0$ and the first excited state $^7F_1$ is not large as compared to the thermal energy. This gives rise to a unique temperature ($T$) independent magnetic susceptibility at low $T$, which is known as Van Vleck paramagnetism (PM) [6]. In general physics terms, Van Vleck PM is defined as the temperature independent positive contribution to the magnetic susceptibility which takes into account the effect of higher energy levels via Zeeman interaction. For rare-earth systems, the energy levels are formed due to the spin-orbit (SO) interaction $\lambda \mathbf{L.S}$ (L and S are total orbital and spin angular momentum, respectively), following the Russel-Saunders coupling scheme [3]. Here, $\lambda$ defines the coupling constant and is equivalent to the energy difference between the ground state and the first excited state. In most of rare earth systems, this energy difference is large compared to $k_BT$ and thus, only the lowest energy state is considered to explain the physical properties of a system. In contrast, for $Eu^{3+}$ based compounds, the excited states are also expected to play a role in determining the physical properties; as the separation energy is comparable to the thermal energy. Also, the value of $\lambda$ determines the $T$ regime in which magnetic susceptibility appears as a plateau, as well as, its magnitude. The value of $\lambda$ is reported to be around 490 K for $EuF_3$ [3], 471 K for $EuBO_3$ [3], 460 K for $Eu_2O_3$ [1] and 458 K for EuOOH [5]. Further, beyond the plateau, most of these compounds show an upturn in the susceptibility at lower temperatures [2, 3, and 4]. This feature was attributed to the presence of magnetic species like $Eu^{2+}$. Ideally, $J = 0$ systems should exhibit zero effective moment ($\mu_{eff}$). But, the presence of closely separated magnetic first excited state ($J = 1$) results in appreciable magnetic susceptibility in $Eu^{3+}$ based materials.

Here, we introduce another Eu based compound, $EuVO_4$, where the emergence of non-zero $\mu_{eff}$ is investigated via the means of different experimental probes. This compound belongs to the family of rare earth orthovanadates $RVO_4$ and the members of this series are reported to exhibit interesting optical and magnetic properties due to the presence of indirect super-exchange magnetic interaction and 4$f$ electron-phonon coupling [7-11]. Recently, this series has again gained a lot of interest as some of the compounds of this series exhibits exotic magnetic properties. In TmVO$_4$, magnetic field tuned ferroquadrupolar quantum phase transition is reported [12, 13, 14] whereas, field tuned quantum criticality is observed in DyVO$_4$ [15]. Along with this, two other members of this series HoVO$_4$ [16] and CeVO$_4$ [17] are reported to show signatures of quantum spin liquid (QSL) state and effective spin – ½ state at low temperatures, respectively. This motivated us to investigate EuVO$_4$, which had been earlier explored in terms of optical, electrochemical and morphological properties [18 – 20]. Along with this, magnetic properties of EuVO$_4$ had also been reported, but the study was limited up to 70 K and was mainly focussed on the anisotropy of the system [21, 22]. However, the magnetic properties below 70 K and the effect of externally applied field on the ground state of this orthovanadate have not been investigated.

Hence, in this work, we report the structural and magnetic studies on $J = 0$ rare earth orthovanadate EuVO$_4$. The $T$ dependent DC susceptibility ($\chi_{DC}$) is manifested by three different features. In the high $T$ region, the $\chi_{DC}$ exhibits a linear dependence on $T$ which is attributed to the crystalline electric field of tetragonal phase. This is followed by a plateau like region and further by a Curie-Weiss like behaviour, as $T$ is lowered. The plateau-like feature is ascribed to Van Vleck PM, while, the low $T$ feature is believed to arise due to the Eu$^{2+}$ spins, giving rise to non-zero $\mu_{eff}$. The susceptibility calculations using modified Van Vleck PM theory matches very well with the experimental data in the entire measured temperature range. It also indicates that $\lambda$ between $J = 0$ and magnetic $J = 1$ states decreases as the magnetic field ($H$) increases, while the magnetic interactions among Eu$^{2+}$ ions strengthen. This reduction in $\lambda$ induces magnetic contribution to the non-magnetic ground state. Thus, from our analysis, it is believed that the admixing of the ground state and the first excited state give rise to an enhanced effective moment in EuVO$_4$ in the presence of external magnetic field.

## 2. Methods:

Polycrystalline sample of EuVO$_4$ was synthesized using the conventional solid state reaction method by taking the high purity Eu$_2$O$_3$ and VO$_2$ from Sigma Aldrich. The preparation method is described in detail in [15]. Room temperature powder X-ray diffraction (XRD) data was collected in the range (10°- 90°) using Rigaku diffraction with Cu K$\alpha$ ($\lambda$ = 1.54) radiation. Rietveld method was used to refine the crystal structure using FullProf Suite software. X-ray photoelectron spectra (XPS) was obtained using a monochromatic Al Kα (1486.6 eV) X-ray source with an energy resolution of 400 meV and Scienta analyser (R3000). The sample surface was cleaned in situ by scraping with a diamond file until sample had the minimum O 1s signal as measured by Al Kα x-ray. The binding energy was calibrated by measuring the Fermi energy of Ag in electrical contact with the sample. The base pressure during the measurement was $5\times 10^{-10}$ mbar. Magnetic field and temperature dependent magnetic measurements in the temperature range 1.8-300 K were performed using the Magnetic Property Measurement System (MPMS) from Quantum Design, USA.

## 3. Results and Discussion:

### 3.1 Crystal structure

Fig. 1 depicts the Rietveld refined powder X-ray diffraction (XRD) data obtained at room temperature. It shows that $EuVO_4$ crystallizes in single phase with tetragonal crystal structure having space group $I4_1/amd$ (No. 141). The refinement parameters $R_p$, $R_{wp}$ and $R_{exp}$ are found to be 10.5 %, 12.6 % and 13.1 %, respectively, with a goodness of fit (GOF) ~ 2.09. The obtained lattice parameters a = b = (7.236 ± 0.003) Å, c = (6.365 ± 0.002) Å and volume (V) = 333.27 Å$^3$ matches very well with previous reports [23]. Other parameters like positional coordinates and occupancy, obtained from the refinement, are mentioned in the Table 1. The inset of Fig. 1 shows the crystal structure of $EuVO_4$. It is comprised of $VO_4$ tetrahedron which shares corner and edges with the $EuO_8$ polyhedron. The magnetic interactions in these $RVO_4$ compounds take place through two different super-exchange paths: (Eu – O – Eu) and (Eu – O – V – O – Eu).

### 3.2 X-ray photoemission spectroscopy

In order to study the valence states of $EuVO_4$, we have performed room temperature X-ray photoemission spectroscopy (XPS). Fig. 2 (a, b and c) shows the core level XPS spectra for Eu 3d, V 2p and O 1s, respectively. Tougaard method is used to subtract the inelastic background [24] and the curves are fitted using the Voigt function. The solid red and green lines in the figure represent the resultant fit and individual peak fits, respectively. Fig. 2 (a) shows the XPS spectra of Eu 3d. It reveals the presence of two major peaks, along with two minor peaks (shown by the black arrows). The two major peaks at 1134.8 eV and 1164.7 eV corresponds to $Eu^{3+}$ $3d_{5/2}$ and $3d_{3/2}$, respectively [25, 26]. While, the other two small peaks at 1125 eV and 1156.1 eV are ascribed to the presence of $Eu^{2+}$ $3d_{5/2}$ and $3d_{3/2}$, respectively [25, 26]. These obtained energy values are slightly different from that reported for $Eu^{3+}$ in $Eu_2O_3$ and $Eu^{2+}$ in $EuCl_2$. However, the spin-orbit splitting energy between the $3d_{5/2}$ and $3d_{3/2}$ states for $Eu^{3+}$ (= 29.9 eV) and $Eu^{2+}$ (= 31.1 eV) is in agreement with the reported splitting in $Eu_2O_3$ and $EuCl_2$ (i.e., 29.9 eV for $Eu^{3+}$ in latter case and 30.1 eV for $Eu^{2+}$ in the former case) [26]. In other Eu-based systems, the presence of these $Eu^{2+}$ species are reported to give rise to a Curie-like increment in the magnetic susceptibility at low temperatures [2 – 4]. This presence of $Eu^{2+}$ spins in Eu-based intermetallic is not unusual and arises due to intermediate valence between $Eu^{3+}$ ($4f^6$) and $Eu^{2+}$ ($4f^7$) electronic configurations [3]. This signifies the presence of intrinsic hybridization between $Eu^{2+}$ and $Eu^{3+}$ ions. In ionic compounds, like $Eu_2O_3$ and $EuF_3$, the possibility of hybridization between these two ions is very less. However, these compounds still show signatures of $Eu^{2+}$ ions and the concentration of these ions do not exceed 0.01 % in both the compounds [3]. This implies the presence of very weak intrinsic hybridization between the two configurations. In our system, this concentration (calculated from magnetic susceptibility) is found to be around 0.02 %. Further, the fitted XPS spectrum for V 2p is shown in the Fig. 2 (b). The spectra reveal the presence of two different oxidation states: $V^{5+}$ and $V^{4+}$. In the spectra, the two peaks at 517.7 eV and 525 eV corresponds to $V^{5+}$ $2p_{3/2}$ and $2p_{1/2}$, respectively. While, the other two small peaks at 515.9 eV and 523.1 eV corresponds to the presence of $V^{4+}$ $2p_{3/2}$ and $2p_{1/2}$, respectively [25, 27, and 28]. Earlier reports suggest that $V^{4+}$ may be induced by the photoemission (i.e., due to X-ray irradiations in the XPS) [25]. The induction of $V^{4+}$ due to X-ray irradiations had been reported in the reference [27]. In this report, the XPS study on $V_2O_5$ (a pure $V^{5+}$-based compound) reveal that on increasing the irradiation time of the XPS spectra, a contribution on the low energy side of V $2p_{3/2}$ arises which is ascribed to the presence of $V^{4+}$ species on the surface. Thus, the observed $V^{4+}$ peaks in $EuVO_4$ are believed to arise due to the irradiation effect of the X-ray in the XPS. Fig. 2 (c) depicts the fitted core XPS spectra of O 1s and reveals the presence of only one peak centred at 530.5 eV which corresponds to the lattice oxygen (i.e., $O^{2-}$ of $EuVO_4$). The observed behaviour is in contradiction to the previously reported XPS study on $EuVO_4$ [23], which reports the presence of another peak at higher energy attributed to the presence of OH group. However, the absence of any such peak in our spectra suggests that our sample is formed in single phase and any kind of impurity oxide is absent.

### 3.3 Magnetic properties

Temperature dependence of DC magnetic susceptibility is measured under the zero-field cooling (ZFC) protocol at various applied magnetic fields up to 70 kOe in the range of 1.8 – 300 K. The diamagnetic contribution to the $\chi_{DC}$ is subtracted from the experimental data, as reported by Landolt-Bornstein [29]. In order to calculate the diamagnetic susceptibility of EuVO$_4$, Pascal's additive law is considered. Fig. 3 (a) depicts the $\chi_{DC}$ vs $T$ curve at 100 Oe, with a solid red line (discussed later). The plot can be divided into three different regions, shown by three different shades. In region I, $\chi_{DC}$ shows a linear increment on decreasing the temperature. This behaviour is in contradiction to the theoretically expected behaviour, as ideally Eu$^{3+}$ ($J$ = 0) should possess non-magnetic ground state. But the crystal field environment created by the surrounding non-magnetic oxygen ligands plays an important role in determining the magnetic properties at higher temperatures. As mentioned in section 3.1, EuVO$_4$ crystallizes in tetragonal structure at room temperature in which the Eu atoms form a polyhedron with the nearest oxygen atoms. These Eu$^{3+}$ ions are surrounded by eight oxygen ions and create tetragonal crystal electric field (CEF) around the Eu ion with D$_{2d}$ symmetry. As mentioned in the introduction section, the spin orbit coupling in Eu$^{3+}$ ions result in non-magnetic ground state ($^7$F$_0$) and a first excited magnetic state ($^7$F$_1$) with separation $\lambda$. This $\lambda$ is not large enough as compared to k$_B$T and hence, the first excited state also gets affected by the CEF. Thus, the magnetic properties at high temperatures can be ascribed to the first excited state as well. In our system, the tetragonal CEF splits the $J$ = 1 state into one singlet and one doublet. Hence, due to this splitting, in spite of exhibiting non-magnetic ground state, Eu$^{3+}$ shows appreciable $\chi_{DC}$ at high temperatures and results in linear $T$ dependence. Further, in order to investigate this feature, we have plotted the $T$ dependent inverse DC susceptibility ($\chi_{DC}^{-1}$) at 100 Oe in the Fig. 3 (b). The solid red solid line shows the fitting to the Curie-Weiss (CW) law of the form $\chi_{DC}$ = C/(T-$\theta_{CW}$). The parameters from the fitting, Curie constant (C) and CW temperature ($\theta_{CW}$) are found to be (3.51$\pm$ 0.12) emu/mol-Oe-K and – (461.97 $\pm$ 5.40) K, respectively. The extracted value of $\mu_{eff}$ from C is ~ 5.37 $\mu_B$. This value is quite unusual as Eu$^{3+}$ having ground state $^7$F$_0$ should have zero $\mu_{eff}$. Further, the large negative value of $\theta_{CW}$ implies the presence of strong AFM interactions which is quite anomalous. This is because Eu$^{3+}$ ions are non-magnetic and ideally should not have any kind of interactions among themselves. This signifies that the conventional CW law cannot be used to describe the magnetic properties of EuVO$_4$. Further, these observed unusual parameters indicate towards the significant contribution of the splitted $J$ = 1 state in determining the magnetic properties of EuVO$_4$ at higher $T$. Region II: As $T$ is further lowered, $\chi_{DC}$ attains temperature independent (TI) plateau like region which is a characteristic of Van Vleck PM [6]. As the $T$ is further decreased, this region vanishes and an upturn in $\chi_{DC}$ is observed. Below this upturn, in region

III, the $\chi_{DC}$ curve follows Curie-Weiss like behaviour. At low temperatures, the magnetic properties are believed to be solely dependent on the ground state as the thermal energy is not enough to stimulate the excitation between the ground state and excited states. In our systems, as the ground state is non-magnetic, the observed behaviour in $\chi_{DC}$ at low temperatures is quite unusual and the effective moment should ideally be zero. As per earlier reports on $Eu^{3+}$-based systems, this kind of behaviour is believed to originate due to the presence of magnetic $Eu^{2+}$ ions [2 – 4]. Thus, the low temperature ground state of $EuVO_4$ depends on the crystal field scheme of minority $Eu^{2+}$ ions.

Further, in order to understand the observed three different regions in $\chi_{DC}$, we have used Van Vleck theory of PM, discussed in the references [3, 5]. The Van Vleck paramagnetic susceptibility ($\chi_M$) is expressed as:

$$\chi_M = N \frac{\sum(2J+1)e^{-E_J/k_BT}\left\{\frac{g^2\mu_B^2 J(J+1)}{3k_BT} + \alpha_J\right\}}{\sum(2J+1)e^{-E_J/k_BT}} \ldots\ldots (1)$$

where $N$ is the number of atoms, $J$ is the total angular momentum, $k_B$ is the Boltzmann's constant, $g$ is the lande $g$ – factor, $\mu_B$ is the Bohr magneton, $\alpha_J$ is the Van Vleck paramagnetic term and $E_J$ defines the energy corresponding to each state. The expressions for the terms $g$, $E_J$ and $\alpha_J$ are described in the reference [3, 5]. The parameter $\alpha_J$ induces a shift in the ground state energy and mixes the states $J$ and $J \pm 1$. The total angular momentum ($J$) is related to $L$ and $S$ which yields the spin-orbit interaction energy $\lambda L.S$. The value of $\lambda$ is believed to determine the contribution of excited states in determining the ground state of a system. In $Eu^{3+}$-based systems, the effect of excited states is also considered in analysing the magnetic susceptibility and thus, the equation for $\chi_M$ becomes [3]:

$$\chi'_M = \frac{N\mu_B^2}{Z}\frac{A}{3\lambda} \ldots\ldots (2)$$

where, the term $A$ is related to the energy separation $\lambda$ and $Z$ is the partition function and is also related to $\lambda$. The expressions for both these terms are given in the reference [3] which contains only one variable $\lambda$. These are given as:

$$Z = 1 - 3e^{-\lambda/k_BT} + 5e^{-3\lambda/k_BT} + 7e^{-6\lambda/k_BT} + 9e^{-10\lambda/k_BT} + 11e^{-15\lambda/k_BT} + 13e^{-21\lambda/k_BT} \ldots\ldots (3)$$

$$A = 24 + \left(13.5\frac{\lambda}{k_B T} - 1.5\right)e^{-\lambda/k_B T} + \left(67.5\frac{\lambda}{k_B T} - 2.5\right)e^{-3\lambda/k_B T} + \left(189\frac{\lambda}{k_B T} - 3.5\right)e^{-6\lambda/k_B T} + \left(405\frac{\lambda}{k_B T} - 4.5\right)e^{-10\lambda/k_B T} + \left(742.5\frac{\lambda}{k_B T} - 5.5\right)e^{-15\lambda/k_B T} + \left(1228.5\frac{\lambda}{k_B T} - 6.5\right)e^{-21\lambda/k_B T} \ldots\ldots (4)$$

The value of $\lambda$ can be uniquely calculated by comparison with the experimental $T$ dependent susceptibility data. Its value determines the magnitude of the TI susceptibility and the temperature below which susceptibility attains the plateau region. Larger the value of $\lambda$, higher will be the temperature below which plateau like region originates and lower will be the magnitude. The magnitude of the plateau region signifies the amount of hybridization between the excited states and the non-magnetic ground state. The solid red line in Fig. 3 (a) shows the calculated susceptibility from the equation (2). The fitted curve matches very well with the experimental data up to the $T$ at which upturn is observed with the parameter $\lambda = (560.09 \pm 0.88)$ K. Below this temperature, a clear deviation is observed which implies that at low temperatures, some extra contribution to the susceptibility dominates.

Further, as mentioned above, the $T$ dependent $\chi_{DC}$ curve exhibits Curie-Weiss like behaviour at low temperatures (region III) which is ascribed to the presence of $Eu^{2+}$ spins. In order to consider this low temperature behaviour, we have modified the equation (2) by considering the Curie-Weiss term. The modified expression is given as:

$$\chi_{DC} = \frac{N\mu_B^2}{Z}\frac{A}{3\lambda} + \frac{C}{T - \theta_{CW}} \ldots\ldots (5)$$

Here, $C$ is the Curie constant and $\theta_{CW}$ is the Curie-Weiss temperature. The second term in equation (3) describes the parameters for the interaction among the $Eu^{2+}$ spins. Fig. 4 shows the $T$ response of $\chi_{DC}$ in the presence of different external $H$ up to 70 kOe. The solid open circles in the Fig. 4 (a) shows the experimental data at 100 Oe along with the solid red line showing the fit to the equation (5). The obtained parameters from the fit are given as: $\lambda = (565.11 \pm 1.01)$ K, $C = (0.00165 \pm 0.00001)$ emu/mol-Oe-K and $\theta_{CW} = -(2.23 \pm 0.19)$ K. The obtained value of $\lambda$ is significantly large than the other $Eu^{3+}$ based systems [1, 3 and 5]. It implies larger separation between the ground state and first excited state and hence, results in the higher value of temperature corresponding to the onset of plateau region. Further, the negative value of $\theta_{CW}$ signifies the presence of antiferromagnetic (AFM) interactions among the $Eu^{2+}$ spins. The $\mu_{eff}$ calculated form the value of $C$ is found to be $(0.115 \pm 0.002)$ $\mu_B/Eu^{2+}$ and is very less compared to the theoretically calculated value of moment for $Eu^{2+}$ state.

Along with this, we have also carried out the estimation of the Curie contribution (observed at low temperatures) to the experimental $\chi_{DC}$. The concentration of $Eu^{2+}$ ions is evaluated to be around (0.021 ± 0.001) % and is found to be in range of that evaluated for other Eu compounds [1, 3]. Further, in order to investigate the effect of applied $H$, we have plotted $\chi_{DC}$ vs $T$ curves at various applied $H$ in the Fig. 4. The curves show quite unusual behaviour. As noted from the Fig. 4 (a), the 10 kOe curve lies below the 30 kOe curve, while the 50 and 70 kOe curves follow almost identical path (from the Fig. 4 (b)). For more clear understanding, we have fitted the $\chi_{DC}$ data with the equation (5) which is shown by the red solid lines in the Fig. 4. Additionally, the inset of Fig. 4 (b) represents the theoretically calculated susceptibility using the obtained parameters from the fitting and is discussed in the next paragraph.

The parameters obtained from the fitting as a function of $H$ are shown in the Fig. 5. Fig. 5 (a) depicts that $\lambda$ exhibits a continuous decrement on increasing $H$. $\lambda$ signifies the magnitude of TI susceptibility and in order to understand this, we have calculated the TI susceptibility using equation (2) corresponding to various $\lambda$ obtained at different $H$. From the inset Fig. 4 (b), it is clearly visible that as $H$ increases (i.e., $\lambda$ decreases), the magnitude of TI susceptibility increases. This is in accordance with the theoretical studies. Further, Fig. 5 (b, c) represents the $H$ variation of $\mu_{eff}$ and $\theta_{CW}$ which shows quite anomalous behaviour. These parameters first increase up to 50 kOe and then decreases. This implies that the $\chi_{DC}$ is quite sensitive to the external $H$ and needs further attention via some microscopic probes. Also, the increment in $\mu_{eff}$ with $H$ hints towards some additional contribution at higher $H$. The decrement in $\lambda$ implies that the energy separation between the $J = 0$ and $J = 1$ decreases. As the first excited state is a magnetic state, the reduction in $\lambda$ induces magnetic contribution to the non-magnetic ground state in the presence of $H$. Further, as seen from the figure, $\theta_{CW}$ starts becoming more and more negative with the increasing $H$. This implies the external $H$ perturbs the magnetic interactions and thus, the AFM correlations among the $Eu^{2+}$ spins become stronger. Thus, from our analysis, it is inferred that at low field, the presence of $Eu^{2+}$ result in $\mu_{eff}$ at low temperatures. As $H$ is increased, the energy levels are altered. This results in a reduced separation between the non-magnetic ground state and excited states, thereby, inducing a magnetic contribution to the ground state. As a result, an increased moment is noted at low temperatures in this compound, on application of external fields.

## 4. Conclusion:

Through our comprehensive experimental studies, it is concluded that $T$ dependent $\chi_{DC}$ is manifested by three different regions. The high $T$ region is attributed to tetragonal crystal field effect, followed by a $TI$ plateau like region which is ascribed to Van Vleck PM. It is followed by Curie-Weiss like behaviour at low $T$. In this region, the ground state is solely determined by $J = 0$ states and the presence of $Eu^{2+}$ ions give rise to the observed trend. The magnetic susceptibility obtained from the modified Van Vleck theory agrees very well with the experimental data. Our results indicate that $\lambda$ decreases as $H$ increases. On the other hand, the AFM interactions among $Eu^{2+}$ become more dominating. The reduced $\lambda$ induces magnetic contribution to the non-magnetic ground state. Thus, in the presence of $H$, the reduced $\lambda$ give rise to an enhanced effective moment. Our studies indicate towards the intricate role of first excited state in determining the magnetic state of $EuVO_4$, in both high as well as low temperature regime. Hence, it can be said that the excited states must be treated as an important parameter for unravelling the underlying physics of Eu-based systems.

**ACKNOWLEDGEMENTS**

The authors acknowledge IIT Mandi for experimental and financial support.

**References:**

1) Borovik-Romanov A S, Kreines N M 1956 *Sov. Phys. JETP* **2** 657.
2) Kern S, Raccah P M, Tveten A 1970 *J. Phys. Chem. Solids* **31** 2639.
3) Takikaw Y, Ebisu S, Nagata S 2010 *J. Phys. Chem. Solids* **71** 1592.
4) Petrov D, Angelov B, Lovchinov V 2011 *J. Alloys Compd.* **509** 5038.
5) Samata H, Wada N, Ozawa T C 2015 *Journal of rare earths* Vol. 33, No. 2 177.
6) Van Vleck J H 1932 The Theory of Electric and Magnetic Susceptibilities, *Oxford University Press*, p. 226.
7) Hirano Y, Skanthakumar S, Loong C K, Wakabayashi N and Boatner L A 2002 *Phys. Rev. B* **66** 024424.
8) Suzuki H, Higashino Y and Inoue T 1980 *J Phys. Soc.* Vol. **49**, 3 1187.
9) Kirschbaum K, Martin A, Parrish D A and Pinkerton A A 1999 *J Phys. Condens. Matter* **11** 4483.

10) Becker P J, Dummer G, Kahle H G, Klein L, Muller-Vogt G, Schopper H C 1970 *Phys. Letts.* A **31** 499.

11) Will G and Schafer W 1971 *J. Phys. C: Solid State Phys.* **4** 811.

12) Massat P, Wen J, Jiang J M, Hristov A T, Liu Y, Feigelson R S, Lee Y S, Fernandes R M and Fisher I R 2022 arXiv:2110.03791v1 [cond-mat.str-el].

13) Vinograd I, Shirer K R, Massat P, Wang Z, Kissikov T, Garcia D, Bachmann M D, Horvatic M, Fisher I R and Curro N J 2022 Cond-matt arXiv: 2112.05859v1.

14) Wang Z, Vinograd I, Mei Z, Menegasso P, Garcia D, Massat P, Fisher I R and Curro N J 2021 *Phys Rev* B **104** 205137.

15) Ranaut D and Mukherjee K 2022 *Sci. Rep.* **12** 56.

16) Ranaut D, Shastri S S, Pandey S K and Mukherjee K. 2022 (Under Review).

17) Ranaut D and Mukherjee K 2022 *J. Phys.: Condens. Matter* **34** 315802.

18) Govindarajan D, Johanson F J, Shankar V U, Salethraj M J and Gopalakrishnan R 2021 *J Mater Sci: Mater Electron* **32** 19434.

19) Conet R L, Hansen P C, Leask M J M and Wanklyn B M 1993 *J Phys: Condens. Matter* **5** 573.

20) Vosoughifar M 2017 *J Mater Sci: Mater Electron* **28** 2227.

21) A Nag, D Ghosh and B M Wanklyn 1998 *Solid state communications, Vol. 108, No. 5, 265.*

22) B Bleaney 1995 *Proc. R. Soc. Lond. A* **450** *711.*

23) Li Li-Ping, Li G S, Xue Y F and Inomata H 2001 *Journal of The Electrochemical Society* **148** (9) J45.

24) Tougaard S 1989 *Surf. Sci.* **216** 343.

25) Ambard C, Duee N, Pereir F, Portehault D, Methivier C, Pradier C M, Sanchez C 2016 *Journal of Sol-Gel Science and Technology* **79** 381.

26) Kim D, Kim S C, Bar J S, Kim S, Kim S J and Park J C 2016 *Inorg. Chem.* **55** 8359.

27) Silversmit G, Depla D, Poelman H, Marin G B, Gryse R D 2006 *Surface Science* **600** 3512.

28) Silversmit G, Depla D, Poelman H, Marin G B, Gryse R D 2004 *Journal of Electron Spectroscopy and Related Phenomena* **135** 167.

29) Hellwege K H, Hellwege (Eds.) A M 1986 *Landolt-Bornstein, New Series, GroupII*, vol. **16**, Springer, Berlin, p.402

Table 1: Position coordinates and occupancies, obtained from the Rietveld refinement of the XRD data of EuVO$_4$ taken at room temperature.

| Atom | Wyckoff position | x | Y | z | Occupancy |
|---|---|---|---|---|---|
| Eu | 4a | 0 | 0.75 | 0.125 | 0.887 |
| V | 4b | 0 | 0.25 | 0.375 | 0.957 |
| O | 16h | 0 | 0.0696 | 0.2132 | 3.882 |

**Figures:**

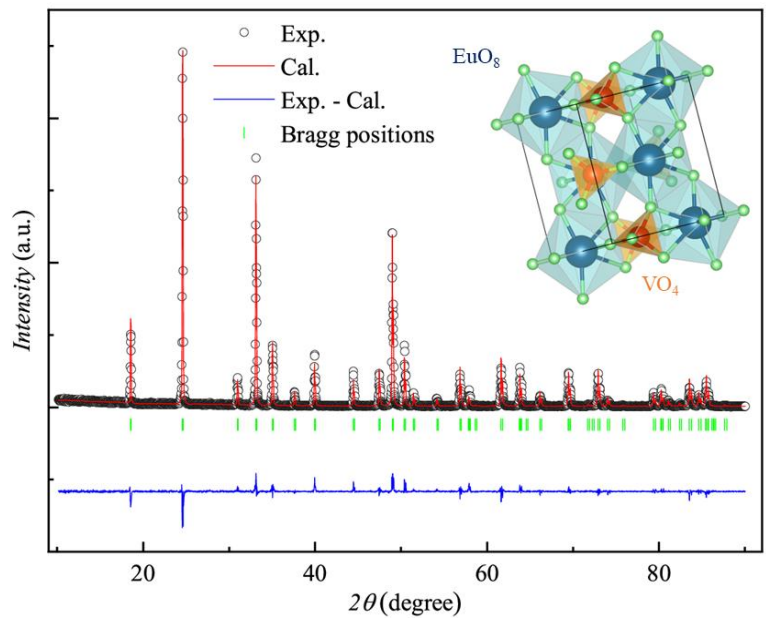

**Fig. 1:** Rietveld refined room temperature XRD data. The black open circles represent the experimental data, while the solid red line indicates the Rietveld refined pattern. The solid blue line and green vertical lines show the difference curve and Bragg positions, respectively. The inset shows the crystal structure of EuVO$_4$ comprised of EuO$_8$ polyhedron and VO$_4$ tetrahedra.

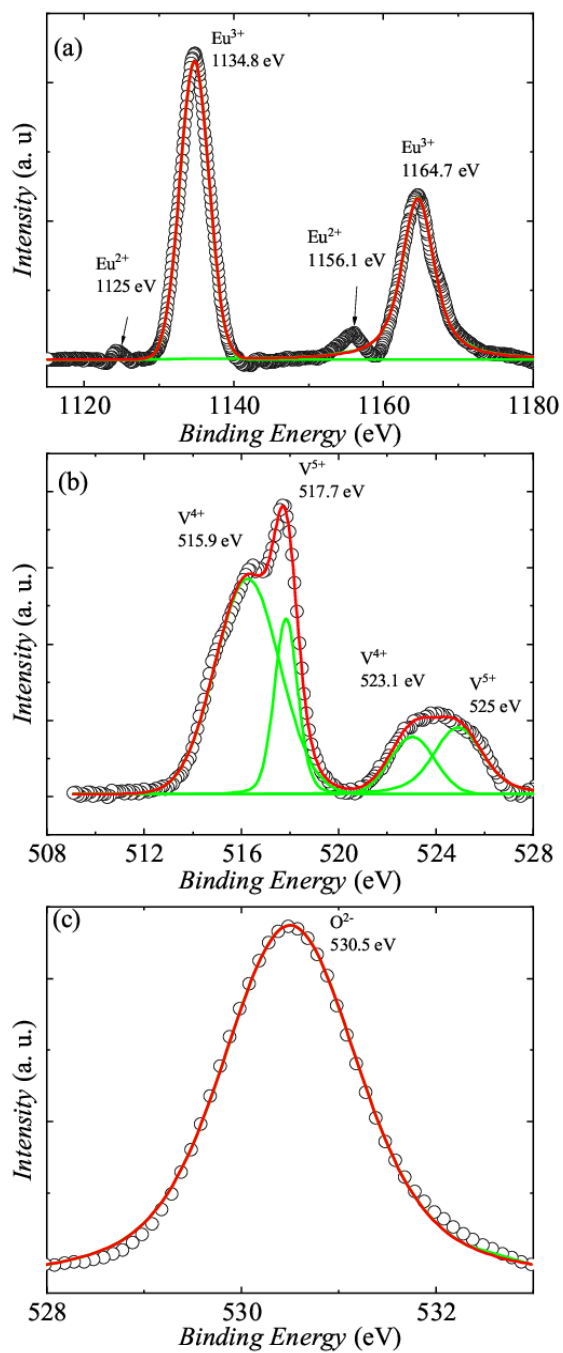

**Fig. 2:** (a, b, c) Room temperature X-ray photoemission spectra (XPS) of the core level of Eu 3*d*, V 2p, and O 1*s*, respectively. The solid red and green lines show the resultant fit and individual peak fittings, respectively.

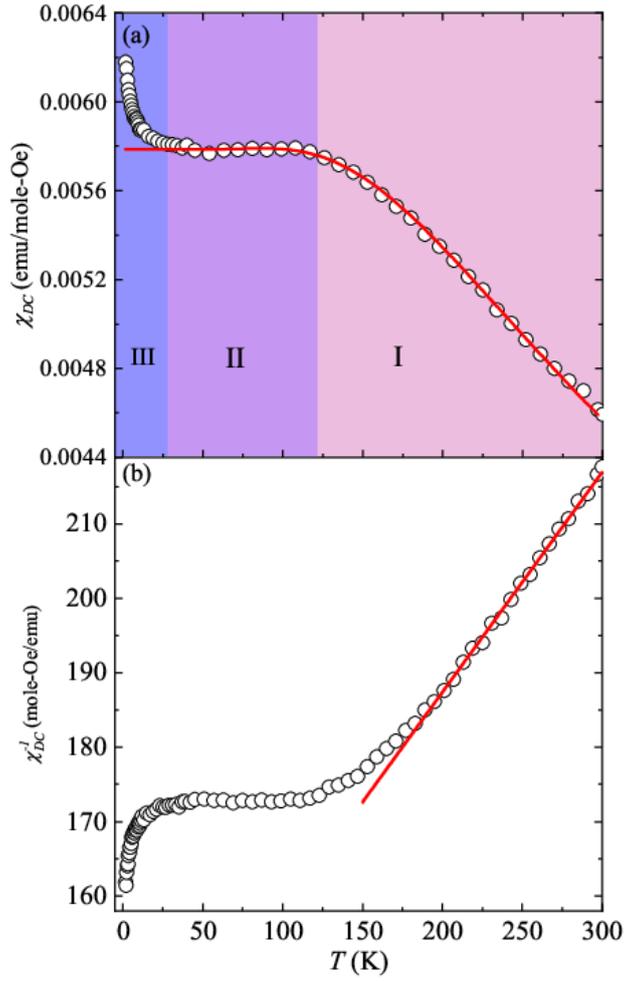

**Fig. 3: (a)** Temperature response of $\chi_{DC}$ measured at 100 Oe. The three shaded regions signify three different features observed in the experimental data. The solid red line shows the fitting to the equation (2). **(b)** $T$ dependent inverse DC susceptibility at 100 Oe, with solid red line showing fit to the Curie-Weiss law.

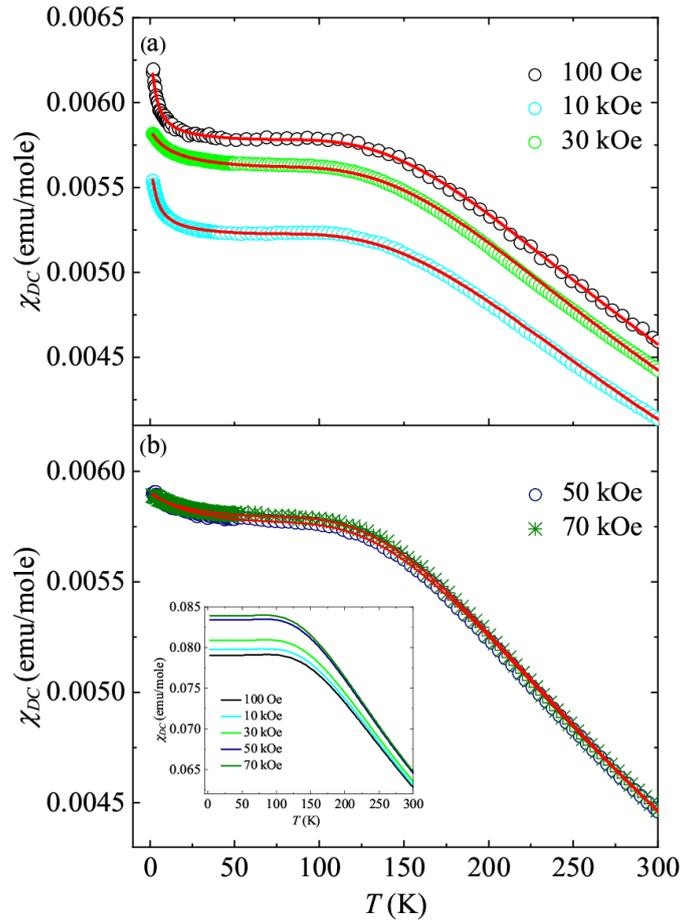

**Fig. 4:** (a, b) $\chi_{DC}$ vs. $T$ curves measured at different externally applied magnetic fields up to 70 kOe. The solid red lines represent the fitted curves corresponding to equation (5). Inset of (b) shows the calculated susceptibility at various applied $H$ using the values of $\lambda$ obtained from the fitting.

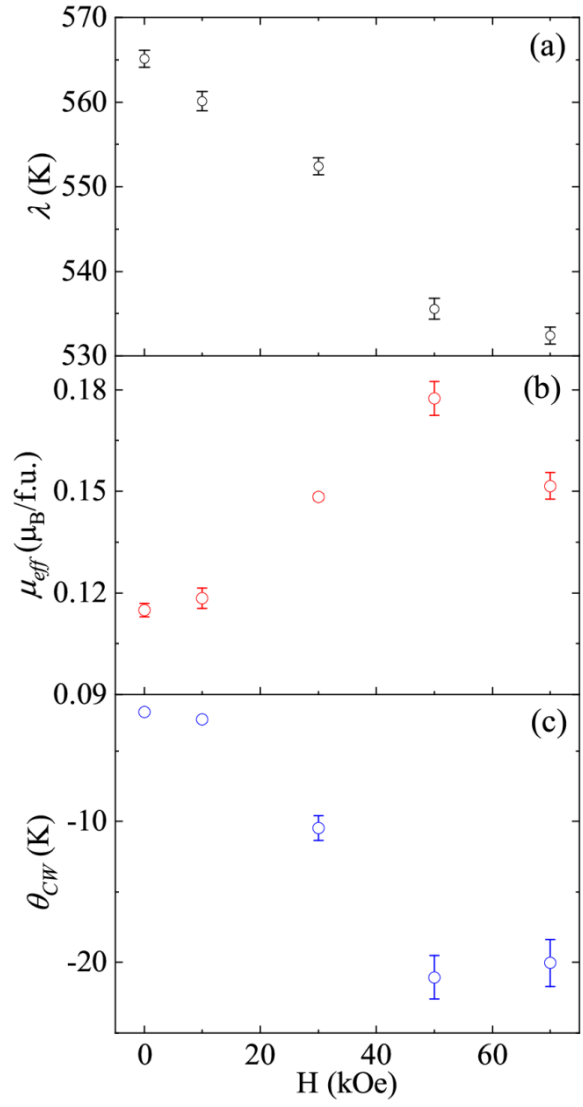

**Fig. 5:** Variation of parameters extracted from the fitting of equation (5) to the experimental $\chi_{DC}$ data as a function of $H$.